\documentclass[aps,pra,twocolumn,showpacs,floatfix]{revtex4-1}
\usepackage{bm}
\usepackage{graphicx} % for PDF output
\usepackage{amsmath}
\usepackage{epstopdf}
\usepackage{subfigure}
\usepackage{array}
\begin{document}
\title{Two-Step Production of Resonant Bose-Einstein Condensate}

\author{M. W. C. Sze$^1$, J. L. Bohn$^1$}
\affiliation{$^1$JILA and Department of Physics, University of Colorado, Boulder, Colorado 80309-0440, USA}

\date{\today}

\begin{abstract}
Producing a substantial and stable resonant Bose-Einstein condensate (BEC) has proven to be a challenging experimental task due to heating and three-body losses that may occur even before the gas comes to thermal equilibrium. In this paper, by considering only two-body correlations, we note that a sudden quench from small to large scattering lengths may not be the best way to prepare a resonant BEC. As an alternative, we propose a two-step scheme that involves an intermediate scattering length, between $0$ and $\infty$, which serves to maximize the transfer probability of $N$ bosons of mass $m$ in a harmonic trap with frequency $\omega$. We find that the intermediate scattering length should be $a\approx3.16N^{-2/3}\sqrt{\hbar/(m\omega)}$ to produce an optimum transition probability of $1.03N^{-1/6}$.

\end{abstract}

%\pacs{}

\maketitle

%%%%%%%%%%%%%%%%%%%%%%%%%%%%%%%%%

\section{Introduction}

Recent experimental efforts have sought to prepare a Bose-Einstein condensate (BEC) of ultracold atoms in a regime where the two-body scattering length $a$ is infinite \cite{Makotyn14_NP, Klauss17_PRL, Eigen17_PRL, Fletcher13_PRL, Fletcher17_Science}. Such a situation is termed a ``resonant'' (or sometimes ``unitary'') BEC.  It represents an unusual situation, inasmuch as the perturbative parameter $na^3$ -- where $n$ is the number density -- is no longer small and the usual field-theoretic ideas struggle to be useful.  This circumstance has hatched a variety of alternative theoretical descriptions, which are in general agreement about the nature of the gas, yet differ in details \cite{Song09_PRL,Cowell02_PRL,Lee10_PRA,Borzov,Diederix11_PRA,FZhou,Yin13_PRA,Ding17_PRA,vanHeugten,Rossi_PRA89,Sykes14_PRA,Smith14_PRL,Sze18_PRA}.  

On the experimental side, producing the resonant BEC is problematic, since the rate of three-body recombination grows rapidly with scattering length.  In the resonant limit, this rate ultimately saturates, but at a large value that ensures the heating and ultimate destruction of the gas within milliseconds.  Under these circumstances, a semblance of the approach to equilibrium can be teased out \cite{Eigen17_PRL,Fletcher17_Science,Eigen18_Nature}, while the loss can be understood as a few-body process incorporating local physics of the gas \cite{Sykes14_PRA,DIncao18_PRL,Bedaque,Braaten06_PhysRep}.  

In order to perform an experiment of this kind at all, the resonant BEC must therefore be produced quickly.  A typical experimental protocol starts with the gas at a small value of scattering length, then rapidly ramps the value of a magnetic field near a Fano-Feshbach resonance so that $a \rightarrow \infty$ within microseconds.  This represents the essentially instantaneous projection of the many-body state at small $a$ onto a collection of many-body states at $a=\infty$.  

While this rapid ramp is essential for defining time zero of the nonequilibrium dynamics, it it not necessarily the best protocol for generating a true resonant BEC.  To see this, at least qualitatively, it is useful to regard the gas within a mean-field-like description.  Consider a gas of $N$ identical bosons, each initially in some single-particle orbital $\phi_a({\bf r})$, corresponding to the small initial scattering length $a$ (the function $\phi_a$ could be the ground state solution to the Hartree-Fock equations for the Bose system, for example).  The many-body wave function is then, to a good approximation,
\begin{eqnarray}
\Psi({\bf r}_1,{\bf r}_2, \dots {\bf r}_N) = \prod_{i=1}^N \phi_a({\bf r}_i).
\end{eqnarray}
Similarly, on resonance each atom can be regarded as belonging to some different orbital wave function $\psi_{\infty}({\bf r})$.  This could be obtained approximately, for example, by performing a Hartree-Fock calculation using a renormalized scattering length $a \propto n^{-1/3}$ \cite{Ding17_PRA,Sze18_PRA, vonStecher07_PRA,FZhou,Lee10_PRA,Song09_PRL}.  Thus, at least up to a certain approximation, the desired resonant BEC is described by
\begin{eqnarray}
\Psi_{\rm res}({\bf r}_1,{\bf r}_2, \dots {\bf r}_N) = \prod_{i=1}^N \psi_{\infty}({\bf r}_i).
\end{eqnarray}
Then the probability that the initial state $\Psi$ produces the resonant BEC states $\Psi_{\rm res}$, assuming that the fast ramp results in a projection, is given by the square of their overlap
\begin{eqnarray}
P = |\langle \Psi | \Psi_{\rm res} \rangle|^2 = 
\left( \left| \int d^3r \phi_a^*({\bf r}) \psi_{\infty}({\bf r})  \right|^{2} \right)^{N}
\end{eqnarray}
Unless each of these overlap integrals is very close to one, the product of $N$ of them will be vanishingly small for typical experimental circumstances with $N>10^3$.   For this reason, it appears that, while the sudden ramp to $a=\infty$ produces an interesting, nonequilibrium gas of strongly-interacting bosons, it is unlikely to generate the desired  resonant BEC.  

In this paper we present an alternative scheme for preparing a resonant BEC, which proceeds in two steps.  In a first step, the scattering length is jumped quickly from a low initial value $a_1\approx0$ to a modest intermediate value $a_2$.  The sudden increase in scattering length causes the BEC to expand; when it reaches the size of the resonant BEC,  the scattering length is suddenly jumped from $a_2$ to $a=\infty$.  For a properly-chosen value of the intermediate scattering length $a_2$, we show that the fraction of atoms converted into a resonant BEC can be non-negligible \cite{Klauss17_Thesis}.  

To describe and carry out calculations of this scheme, we focus on an isotropic, harmonically trapped BEC and employ a coordinate-based representation of the BEC wave function.   This representation presents the BEC as a wave packet subject to an effective potential energy surface (PES) \cite{Bohn98_PRA}, and it has recently been shown to make a reasonable description of the BEC on resonance \cite{Ding17_PRA,Sze18_PRA}.  It presents the dynamics as the time evolution of a wave packet obeying a linear Schr\"{o}dinger equation.  In these terms the two-step process is reminiscent of vibrational wave packet dynamics in molecular physics \cite{Garraway02}. It is also amenable to analytic approximations, which will yield simple estimates for the optimum value of the intermediate scattering length $a_2$, as well as the approximate yield of atoms in the resonant BEC at the end of the two steps.

%%%%%%%%%%%%%%%%%%%%%%%%%%%%

\section{Potential Energy Surfaces, Quench from Non-Interacting to Resonant BEC's}
Here, we summarize the theory, which is detailed in Ref. \cite{Sze18_PRA}.

We exploit a coordinate representation of BEC that is expressed in terms of potential energy surfaces (PES's) analogous to Born-Oppenheimer (B.-O.) curves in molecular physics. We define a single, collective coordinate, the hyperradius $\rho$, which represents the size of the condensate. This hyperradius can be expressed as the root-mean-squared interparticle spacing for any configuration of atoms \cite{Sorensen02_PRA}:
\begin{eqnarray}
\rho^2 =  \frac{ 1 }{ N } \sum_{i<j}^{N} r_{ij}^2 .
\end{eqnarray}
All remaining coordinates, collectively denoted by $\Omega$, span a hypersphere of radius $\rho$ in a $(3N-4)$-dimensional configuration space. If the center of mass coordinates are separated, then the Hamiltonian describing the relative motion is given by \cite{Sorensen02_PRA}
\begin{align}
H_{\rm{rel}} = - \frac{ \hbar^2 }{ 2m } \left[ \frac{ 1 }{ \rho^{3N-4} } \frac{ \partial }{ \partial \rho } \rho^{3N-4} 
\frac{ \partial }{ \partial \rho } - \frac{ {\Lambda}_{N-1}^2 }{ \rho^2 } \right] + \frac{ 1 }{ 2 } m \omega^2 \rho^2,
\label{eq:full_H_hyper}
\end{align}
where $m$ is the atomic mass and $\omega$ is the angular frequency of the isotropic harmonic trap. Thus the kinetic energy has a radial part and an angular part, the latter given in general by the grand angular momentum $\Lambda_{N-1}^2$. Two-body interparticle coordinates are encoded in the angular component. Realistic two-body potentials between atoms \cite{Das04_PRA,Das07_PRA,Chakrabarti08_PRA,Lekala14_PRA}, or boundary conditions with realistic scattering lengths \cite{Sorensen02_PRA,Sorensen03_PRA,Sorensen04_JPB,Sogo05_JPB} may be applied to solve the hyperangular component of $H_{\rm{rel}}$. 

Under the B.-O. approximation, the hyperradius $\rho$ is treated as the slow coordinate. That is, at each value of $\rho$, the Schr\"odinger equation, $H_{\rm{rel}} \Psi(\rho,\Omega) = E_{\rm{rel}} \Psi(\rho,\Omega)$ with $E_{\rm{rel}}$ the energy of the relative motion, is solved in the coordinates $\Omega$ to yield a set of eigenenergies $V_\nu(\rho)$. A coupled set of differential equations is obtained if we expand the wave function $\Psi$ in adiabatic hyperangular basis,
\begin{eqnarray}
\Psi = \rho^{-(3N-4)/2} \sum_{\{ \lambda \}} F_{\{ \lambda \}}(\rho) Y_{\{ \lambda \}}(\rho; \Omega),
\end{eqnarray}
for some set of radial expansion functions $F_{\{ \lambda \}}$, and $Y_{\{ \lambda \}}$ are the eigenstates of $\Lambda_{N-1}^2$. However, by applying the B.-O. approximation we assume that the hyperradial kinetic energy operator does not affect $Y_{\{ \lambda \}}$. Hence the differential couplings can be neglected. For simplicity, we also reduce the collective set of quantum numbers $\{ \lambda \}$ to a single quantum number $\nu$, describing excitations in a single hyperangle $\alpha$, where $\sin \alpha = r_{12}/\left(\sqrt{2} \rho \right)$, which incorporates two-body correlations. Thus, we write the wave function as 
\begin{eqnarray}
\label{eq:wavefxn}
\Psi = \rho^{-(3N-4)/2} F_\nu(\rho) Y_{\nu}(\rho;\alpha).
\end{eqnarray}

Using a single adiabatic function, the Schr\"odinger equation becomes a single ordinary differential equation in $\rho$: 
%\begin{widetext}
\begin{align}
\left[ - \frac{ \hbar^2 }{ 2m } \frac{ d^2 }{ d \rho^2 } + V^{\rm diag}(\rho) + V_\nu(\rho)\right] F_\nu(\rho)
= E_{\rm{rel}} F_\nu(\rho),
\label{eq:adiabatic_equations}
\end{align}
%\end{widetext}
where
\begin{align}
\label{eq:Vdiag}
V^{\rm diag}(\rho) &= \frac{ \hbar^2 }{ 2m } \frac{ (3N-4)(3N-6) }{ 4 \rho^2 }
+ \frac{ 1 }{ 2 } m \omega^2 \rho^2,\\
V_\nu(\rho) &= \frac{ \hbar^2 }{ 2 m \rho^2 }  \langle \nu | \Lambda_{N-1}^2 | \nu \rangle.
\end{align}
$V^{\rm diag}$ is the diagonal potential whose ground state supports the non-interacting condensate wave function. $V_\nu$ represents the interaction potential since the state $| \nu \rangle$ is defined by the boundary conditions based on the scattering length which describes the two-body interaction. Exact calculation of the matrix element $\langle \nu | \Lambda_{N-1}^2 | \nu \rangle$, which involves integration over the entire hypersphere, is not trivial. In this paper, we use the results of Ref. \cite{Sze18_PRA}, where some convenient approximations have been made to obtain a meaningful outcome even for $a=\infty$. Also, we apply only this single matrix element  that is representative of the interaction of the lowest hyperangular state of the condensate with any given $a$. Thus, for any scattering length $a$, 
we find a B.-O. potential $V^a(\rho)= V^{\rm diag}(\rho) + V_\nu(\rho)$ with associated hyperangular wave function $\Phi^a(\rho; \Omega)$.  Vibrational states in the PES $V^a(\rho)$ constitute the radial wave functions $F_n^a(\rho)$, each vibration $n$ describing a breathing mode excited above the ground state condensate with $n=0$. The states relevant to our model are, therefore, defined by the scattering length $a$ and the number of breathing quanta,
\begin{equation}
 |a,n\rangle = \rho^{-(3N-4)/2} F_n^a(\rho) \Phi^a(\rho;\Omega).
\end{equation}
\begin{figure}[!htbp]
\centering
\includegraphics[scale=0.6]{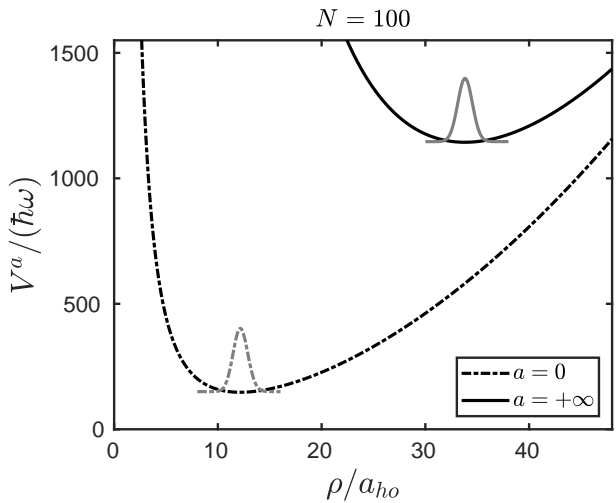}
\caption{\label{fig:Veff_N=100}  The scale of the problem.  Each curve represents an effective potential energy surface for a BEC with $a=0$ (bottom) and $a=\infty$ (top), in our hyperspherical representation. A BEC having $a=0$ (Gaussian centered at $\rho=12.2\, a_{ho}$) has essentially no overlap with a resonant BEC having $a=\infty$ (Gaussian centered at $\rho=33.8\, a_{ho}$).}
\end{figure}

Figure ~\ref{fig:Veff_N=100} shows the B.-O. PES's for the non-interacting ($a=0$) and resonant ($a=\infty$) cases for a gas of $N=100$ atoms. With $a=0$, $V^{0}(\rho) =V^{\rm diag}(\rho)$. This PES, the lowest curve on the left, is exact. The topmost curve on the right is an approximate surface for the resonant limit. Considering only two-body correlations explicitly, this surface is constructed based on Ref. \cite{Sze18_PRA}. For the large $N$ limit, this is given by
\begin{align}
\label{eq:Vinf}
 V^{\infty}(\rho) = \frac{\hbar^2}{2m\rho^2}\left( \frac{9N^2}{4} + 3c_0 N^{8/3} \right) +\frac{1}{2}m\omega^2 \rho^2,
\end{align}
where $c_0\approx2.122$ is a constant determined by the root of some transcendental equation. For realistic values of $N>10^2$, the centrifugal term with $9N^2/4$ can be safely neglected. The ground states of these PES's represent the non-interacting and resonant BEC's. Near their minima \cite{Sze18_PRA},
\begin{align}
\label{eq:rho_0}
 \rho^0 &\underset{N \gg 3}{\approx} \sqrt{\frac{3N}{2}} a_{ho},\\
 \label{eq:rho_inf}
 \rho^\infty &\underset{N \gg 3}{\approx} \left( 3c_0 \right)^{1/4} N^{2/3} a_{ho},
\end{align}
where $a_{ho}=\sqrt{\hbar/(m\omega)}$, we approximate these potentials as harmonic oscillators:
\begin{align}
\label{eq:V0}
V^0(\rho) &\approx  \frac{3N}{2} \hbar \omega + \frac{1}{2}m(2\omega)^2 (\rho-\rho^0)^2, \\
\label{eq:VInf}
V^\infty(\rho) &\approx \left( 3c_0 \right)^{1/2} N^{4/3} + \frac{1}{2}m(2\omega)^2(\rho-\rho^\infty)^2.
\end{align}
In both cases, the excitation frequency of the radial breathing modes considered is exactly twice the trap frequency, $\omega_b=2\omega$. For non-interacting bosons, the energies are well-known and are given by \cite{Smirnov77}
\begin{equation}
 E_{nK}=\hbar \omega \left(2n + K + \frac{3N-3}{2} \right), \qquad n=0,1,2,...
\end{equation}
and $K=0,1,2,...$ is the quantum number associated with the hyperangular component. For the resonant gas, the $2\omega$ frequency was anticipated by symmetry considerations in Refs. \cite{Castin_Physique5,Werner_PRA74}. Without considering three-body or higher order correlations, these references also emphasize that the B.-O. approximation is exact in the $a=\infty$ limit. Corrections beyond the B.-O. approximation arise because the adiabatic wave functions $\Phi$ change from one value of $\rho$ to the next. But this change is only effective if $\rho$ changes significantly on the scale of $a$, i.e., the corrections are of order $\rho/a$ and vanish in the infinite scattering length limit. Therefore, if the atoms could be prepared in the state $F^\infty \Phi^\infty$ that we describe, this state would be stable against non-adiabatic transitions to whatever other states there are that could lead to heating, loss, etc. This stability is likely reduced if we were to include explicit three-body correlations in the wave function.

From the harmonic oscillator nature of the potential curves in Eqs.~(\ref{eq:V0}) and ~(\ref{eq:VInf}), the expected ground state hyperradial wave functions should be Gaussians centered at the minima and with root-mean-squared width of $a_{ho}/\sqrt{2}$:
\begin{align}
\label{eq:F0}
F^0 (\rho) &=\left(\frac{2}{a_{ho}^2 \pi} \right)^{1/4} \exp[-(\rho- \rho^0)^2/a_{ho}^2],\\
\label{eq:FInf}
F^\infty (\rho) &=\left(\frac{2}{a_{ho}^2 \pi} \right)^{1/4} \exp[-(\rho- \rho^\infty)^2/a_{ho}^2],
\end{align}
The unnormalized Gaussian functions $F^0$ and $F^\infty$ for $N=100$ are illustrated as Gaussian-shaped humps at the bottom of the $a=0$ and $a=\infty$ PES's, respectively, in Fig.~\ref{fig:Veff_N=100}. From this picture,  we see that the centers are far away from each other such that quenching the gas suddenly from $a=0$ to $a=\infty$ will yield a low transfer probability. That is, the probability of the atoms landing in the resonant BEC state $F^\infty$, upon a direct quench, is
\begin{align}
 \left|\langle 0,0,| \infty,0\rangle\right|^2 &=\left|\int d\rho F^0(\rho) F^\infty(\rho)\right|^2 \left|\int d\Omega \Phi^0(\Omega) \Phi^\infty(\Omega)\right|^2\nonumber\\
 &\leq \left|\int d\rho F^0(\rho) F^\infty(\rho)\right|^2 \approx \exp{\left( -1.3 N^{4/3} \right)},
\end{align}
which is negligible for large $N$.

\section{The Two-Step Scheme}

\subsection{Franck-Condon Factors}
\begin{figure}[!htbp]
\centering
\includegraphics[scale=0.6]{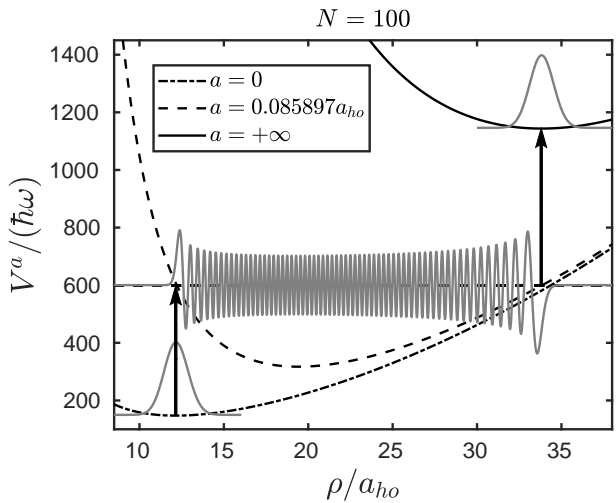}
\caption{\label{fig:TwoStepScheme} The two-step scheme from non-interaction to small $a$ then to resonance.}
\end{figure}
\begin{figure*}[htbp!]
\centering
\begin{subfigure}[]{}
\centering
\includegraphics[width=.3\textwidth]{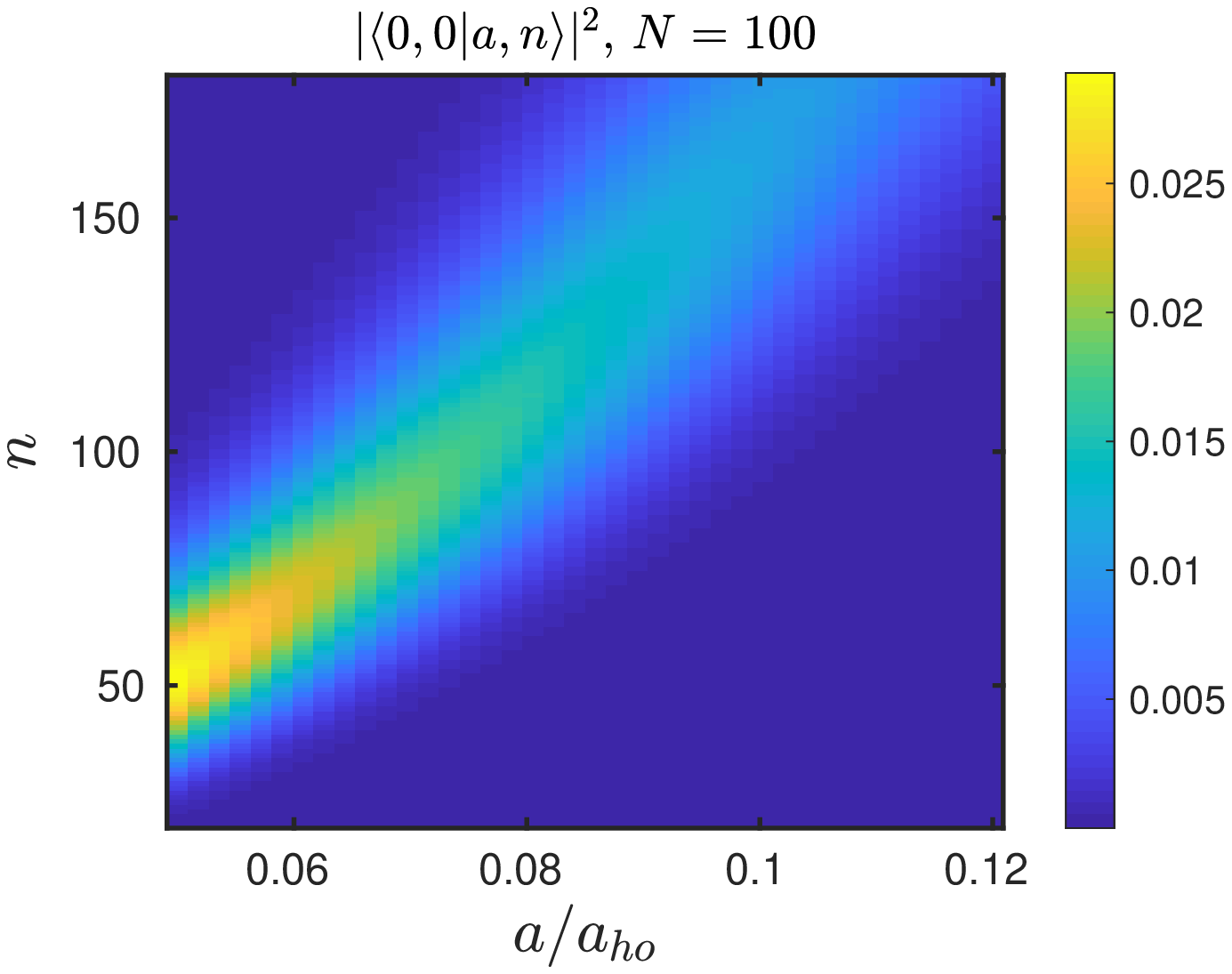}
%\caption{\label{fig:Wavefxn_N=10smallasc}$a=0.0398$, $\rho=3.8$, $\alpha_d=0.1625$}
\end{subfigure}\quad
\begin{subfigure}[]{}
\centering
\includegraphics[width=.3\textwidth]{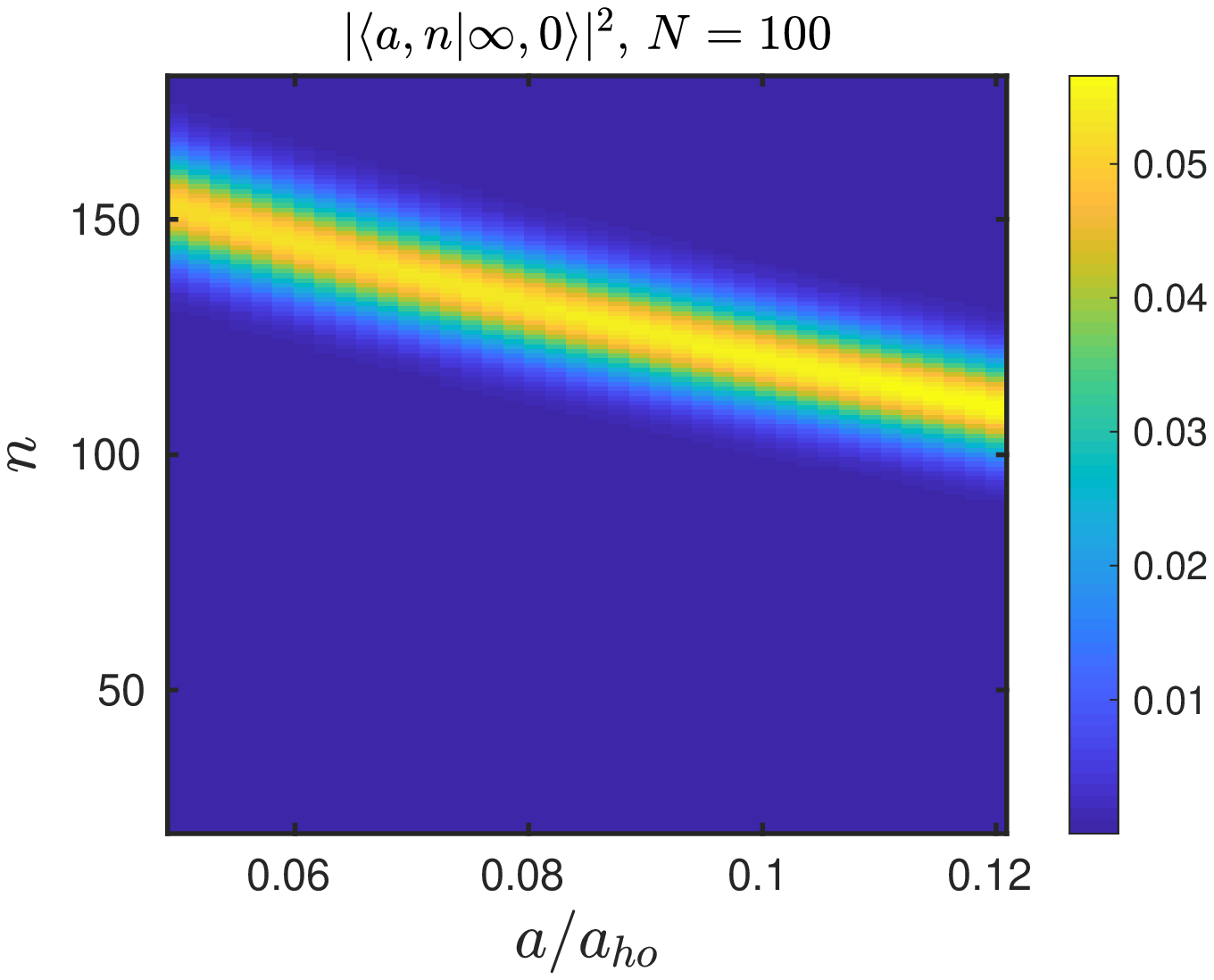}
%\caption{\label{fig:Wavefxn_N=10largeasc}$a=0.0398$, $\rho=3.8$, $\alpha_d=0.1625$}
\end{subfigure}\quad
\begin{subfigure}[]{}
\centering
\includegraphics[width=.3\textwidth]{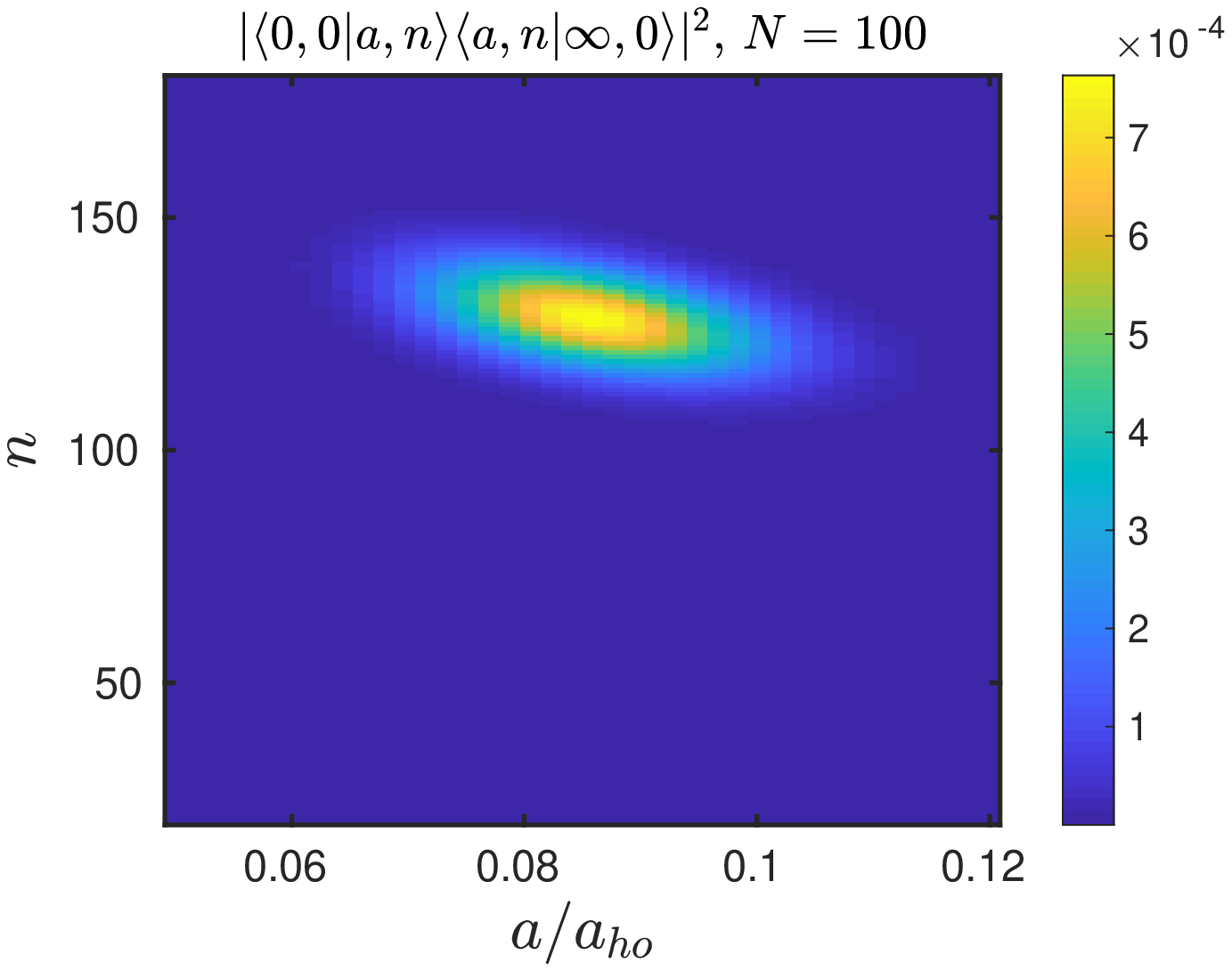}
%\caption{\label{fig:Wavefxn_N=10largeasc}$a=0.0398$, $\rho=3.8$, $\alpha_d=0.1625$}
\end{subfigure}
\caption{\label{fig:TP_N=100} Franck-Condon factors from the (a) non-interacting to intermediate states $|\langle 0,0| a,n \rangle|^2$, (b) intermediate to resonant states $|\langle a,n| \infty,0 \rangle|^2$, and (c) the two-step transition probability  $|\langle 0,0| a,n \rangle\langle a,n| \infty,0 \rangle|^2$ as functions of scattering lengths $a$ and vibrational states $n$. Here, $N=100$. }
\end{figure*}
\begin{figure}[!htbp]
\centering
\includegraphics[scale=0.5]{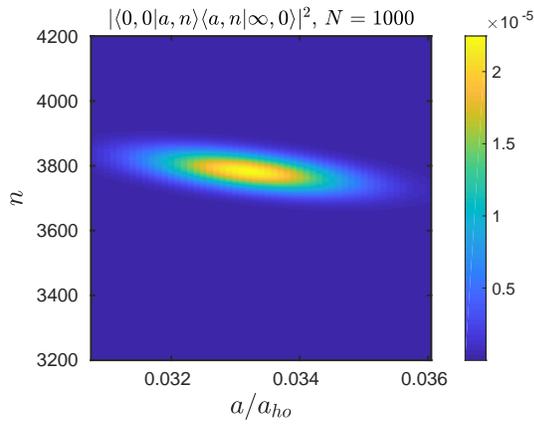}
\caption{\label{fig:TP_N=1000} The two-step transition probability distribution  $|\langle 0,0| a,n \rangle\langle a,n| \infty,0 \rangle|^2$ as a function of scattering lengths $a$ and vibrational states $n$ for $N=1000$.}
\end{figure}

The tiny overlap between $F^0$ and $F^\infty$ suggests that direct projection from $a=0$ to $a=\infty$ will not yield a good amount of resonant BEC. We then seek an intermediate state with finite, nonzero value of $a$. Such a PES, $V^a(\rho)$, is shown as the intermediate curve in Fig.~\ref{fig:TwoStepScheme}. A good candidate for $V^a(\rho)$ is one that supports a set of  vibrational excitations $n$ so that $|a,n\rangle$ has a good overlap with both the $|0,0\rangle$ and $|\infty,0\rangle$ states as shown in Fig.~\ref{fig:TwoStepScheme}. Real BEC experiments have $N>10^4$ atoms. In Fig.~\ref{fig:TwoStepScheme}, we use $N=100$ as an illustrative example. For larger $N$, $\rho^0$ and $\rho^\infty$ grow farther apart. One then needs to use higher vibrational states (with larger number of nodes $n$) to optimize the overlaps $\langle 0,0| a,n \rangle$ and $\langle a,n| \infty,0 \rangle$. These squared overlaps $|\langle 0,0| a,n \rangle|^2$ and $|\langle a,n| \infty,0 \rangle|^2$ are called Franck-Condon (FC) factors. Numerical calculations of these FC factors $|\langle 0,0| a,n \rangle|^2$ and $|\langle a,n| \infty,0 \rangle|^2$, and $|\langle 0,0| a,n \rangle\langle a,n| \infty,0 \rangle|^2$ are reflected as color-map plots in Fig.~\ref{fig:TP_N=100} for $N=100$; the x-axis is the scattering length, y-axis the vibrational state $n$, and the color indicates the transition probability. In general, for the first step from the non-interacting to the intermediate, the optimum transition occurs when $a$ is small and for low $n$ states, decreasing quickly with increasing $a$ and $n$ as shown in Fig.~\ref{fig:TP_N=100}(a).  For the second step from intermediate to final, the transition is optimum when $a$ and $n$ are larger, and diminishes slowly with decreasing $a$ and increasing $n$ as in Fig.~\ref{fig:TP_N=100}(b). These two steps cannot be individually at their maxima under the same conditions. However, the best overall yield occurs when $a$ is still small relative to the oscillator length and for higher vibrational states. This is true for any large values of $N$.  Further, the two-step transition probabilities seem to decrease as a function of $N$. See the transition probability for $N=1000$ in Fig.~\ref{fig:TP_N=1000}.
%\begin{widetext}
%\end{widetext}

\subsection{The Optimum Intermediate State}
Since the intermediate state will have a small value of $a$, we can use a perturbative approximate expression for $V^a$. In the limits of perturbative $a\ll a_{ho}$ and large $N$, this is given by
\begin{equation}
\label{eq:Va}
 V^a(\rho) \underset{N\gg3}{\approx} V^0(\rho) + \frac{\hbar^2}{m} d_0 N^{7/2} \frac{a}{\rho^3},
\end{equation}
where $d_0=(3/4)\sqrt{3/\pi}\approx 0.733$. This potential can be well utilized by considering the classical inner and outer turning points, $\rho_{1n}$ and $\rho_{2n}$, of $V^a$ at some particular energy $E_n$ of $F_n$. At high vibrational states $n$, $\rho_{1n}$ and $\rho_{2n}$ can be approximated through
\begin{align}
\label{eq:rho1n}
 V^a(\rho_{1n})&=E _n^a \underset{N\gg3}{\approx} \frac{\hbar^2}{m} d_0 N^{7/2} \frac{a}{\rho_{1n}^3}, \\
 %\text{ or } \rho_{1n} \approx \left(\frac{\hbar^2}{m} d_0\right)^{1/3} N^{7/6} a^{1/3} {E_n^a}^{-1/3}  \\
\label{eq:rho2n}
V^a(\rho_{2n}) &=E_n^a \approx \frac{1}{2} m \omega^2 \rho_{2n}^2,
\end{align}
where $V^a$ is dominated by the interaction term at small $\rho$, and by the trapping potential at large $\rho$. An effective two-step scheme is illustrated in Fig.~\ref{fig:TwoStepScheme}. It is achieved when the inner turning point $\rho_{1n}$ of $V^a$ is near $\rho^0$ and the outer turning point $\rho_{2n}$ is near $\rho^\infty$. Thus, with $\rho_{1n}\approx \rho^0$, $\rho_{2n}\approx \rho^\infty$ and Eqs.~(\ref{eq:rho2n}) and ~(\ref{eq:rho1n}), the state which would give the maximum Franck-Condon overlap is one whose scattering length and energy are
\begin{align}
\label{eq:Intermediate_a}
a^*&\underset{N\gg3}{\approx}\frac{1}{2d_0} \left(\rho^0\right)^3 \left(\rho^\infty \right)^2 N^{-7/2} \frac{1}{a_{ho}^4}\\
\label{eq:Emax}
 E^* &\approx \frac{1}{2} \left(\frac{\rho^\infty}{a_{ho}}\right)^2 \hbar \omega.
\end{align}
Using these approximations for $N=100$, the results are $a^*=0.145 a_{ho}$ and $E^*= 571.2 \hbar\omega$ which are close to the exact calculations of $a^*=0.0859 a_{ho}$ and $E^*=598.9 \hbar \omega$, the latter set of values can be visually estimated through Figs.~\ref{fig:TP_N=100} and~\ref{fig:TwoStepScheme}. Expressions ~(\ref{eq:Intermediate_a}) and~(\ref{eq:Emax}) become better estimates for larger $N$. For $N=1000$, the predicted results are $a^*=0.0316 a_{ho}$ and $E^*=1.26(10^4) \hbar \omega$, and the numerical computations give $a^*=0.0332 a_{ho}$ and $E^*=1.28(10^4) \hbar \omega$.

While this static picture provides overall orientation, it does not describe the dynamics involved.  Roughly, upon the initial projection from $a=0$ to the intermediate value $a^*$, a wave packet is formed at $\rho_{1n}$.  In approximately one half of the trap period, this wave packet propagates to $\rho_{2n}$, giving the condensate its maximum radial extent and preparing it for projection onto the resonant BEC state.  

\subsection{Wave Packet Dynamics}
\begin{figure}[htbp!]
\centering
\begin{subfigure}[]{}
\centering
\includegraphics[width=.45\textwidth]{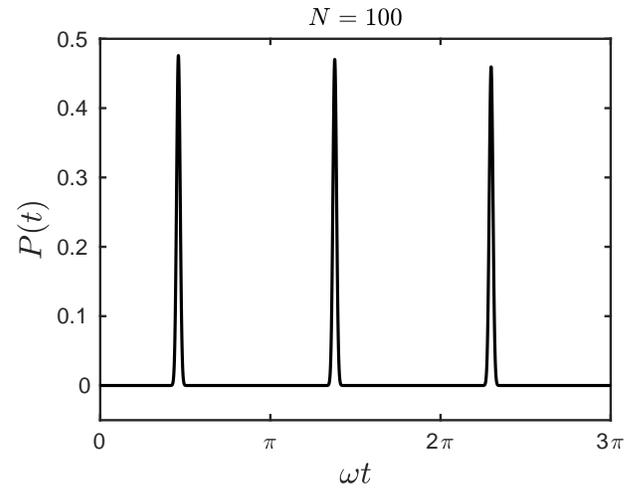}
%\caption{\label{fig:TransProbVsTime_N=100}$a=0.0398$, $\rho=3.8$, $\alpha_d=0.1625$}
\end{subfigure}\qquad\qquad \qquad
\begin{subfigure}[]{}
\centering
\includegraphics[width=.45\textwidth]{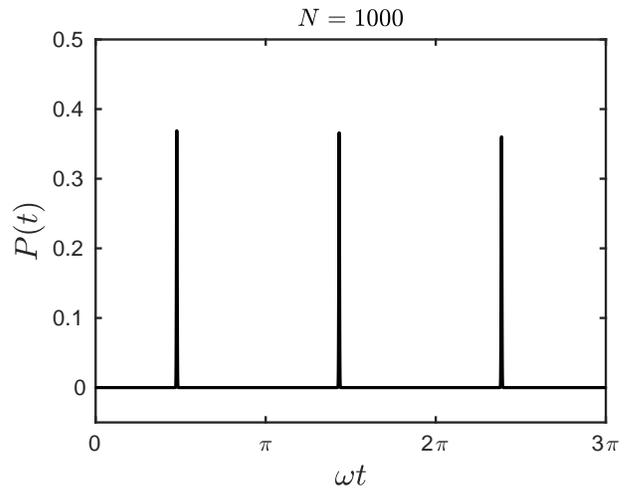}
%\caption{\label{fig:MeanRadiusVsTime_N=100}$a=0.0398$, $\rho=3.8$, $\alpha_d=0.1625$}
\end{subfigure}
\caption{\label{fig:Dynamics} Transfer probability for (a) $N=100$ with $a^*=0.0859 a_{ho}$, and (b) $N=1000$ with $a^*=0.0332 a_{ho}$.}
\end{figure}
\begin{figure}[!htbp]
\centering
\includegraphics[scale=0.58]{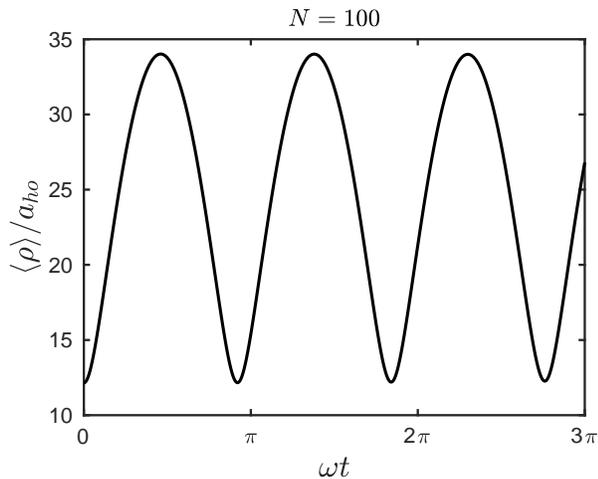}
\caption{\label{fig:MeanRadius_N=100} Mean radius of the BEC versus time for $N=100$ and $a=0.0859 a_{ho}$.}
\end{figure}
To describe the time dynamics, we express the initial state after the first step as a wave packet expanded in the basis of the vibrational states of the intermediate potential
\begin{align}
 |\Psi^a(t)\rangle &= \sum_{n=0}^{\infty} |a,n\rangle \langle a,n| \Psi^a(t=0)\rangle e^{-iE_nt/\hbar}\nonumber\\
 &= \sum_{n=0}^{\infty} |a,n\rangle\langle a,n|0,0 \rangle e^{-iE_n t/\hbar},
\end{align}
where at time $t=0$, $\Psi^a$ is at the ground state of the non-interacting potential with total energy $E\approx 3N\hbar\omega/2$. The probability the projection of the wave packet onto the desired resonant BEC ground state is given by 
\begin{align}
 P(t)=|\langle \infty,0|\Psi^a(t)\rangle|^2,
\end{align}
where
\begin{align}
\label{eq:TransitionAmp}
 \langle \infty,0|\Psi^a(t)\rangle = \sum_{n=0}^{\infty} \langle \infty,0|a,n\rangle\langle a,n|0,0 \rangle  e^{-iE_n t/\hbar}.
\end{align}

After extracting the most appropriate choice for the intermediate $a$, we compute this transition probability at different times with the unitary BEC model found in Ref.\cite{Sze18_PRA} for $N=100$ and $N=1000$. Figures~\ref{fig:Dynamics}(a) and~(b) show that the first maximum transition occurring at around $t_m \approx \pi/(2\omega)$, with $48\%$ transfer probability for $N=100$ and $36\%$ for $N=1000$. It takes about half a period, $T/2$, for the BEC to expand to resonance starting from the left side of the $V^a$; the breathing mode frequency is close to $2\omega$, thus the dwell time is $t_m \approx T/2 = \pi/\omega_b= \pi/(2\omega)$. 

Figure ~\ref{fig:MeanRadius_N=100} shows how the size of the BEC with $N=100$ atoms, expressed in terms of the mean hyperradius $\langle \rho \rangle$, is changing over time. It starts with $\rho=\rho^0$, the size of the non-interacting gas, and reaches $\rho=\rho^\infty$, the size of the resonant BEC, at $t\approx t_m$. The peaks of $P(t)$ and $\langle \rho \rangle$ decrease slowly over time as the wave packet gradually dephases.  It is, therefore, worthwhile to instigate the second projection, to resonance, at time $t=T/2$.

\section{Large $N$ Limit}
In calculating the $P(t)$ numerically, we notice that $P(t_m)$ decreases with $N$. Determining how $P(t_m)$ scales with $N$ is extremely useful. Here, we outline a method to get a good estimate for this scaling. The details are found in the Appendix, and the final result turns out to be simple. 

Using the results from Appendix~\ref{appendix:FC_factors}~and~\ref{appendix:AngularOverlap}, the overlap integrals in Eq.~(\ref{eq:TransitionAmp}) are approximated to be
\begin{align}
 \langle a,n|0,0 \rangle &= \langle F_n^a|F^0 \rangle_\rho \langle \Phi^a|\Phi^0 \rangle_\Omega \nonumber\\
 &\approx F^0(\rho_{1n}) \sqrt{\frac{dE_n}{dn}}\sqrt{\frac{1}{|\partial V^a/\partial \rho|_{\rho_{1n}}}}, \\
  \langle \infty,0 |a,n \rangle &= \langle F^\infty|F_n^a \rangle_\rho\langle \Phi^\infty|\Phi^a \rangle_\Omega \nonumber\\
  &\approx  (-1)^n  F^\infty(\rho_{2n}) \sqrt{\frac{dE_n}{dn}}\sqrt{\frac{1}{|\partial V^a/\partial \rho|_{\rho_{2n}}}},
\end{align}
where $dn/dE_n$ is the density of vibrational states in the intermediate potential. For the hyperangular parts of the wave function we approximate
\begin{align}
\label{eq:PhiaPhi0}
\langle \Phi^a|\Phi^0 \rangle_\Omega &\approx 1 -\frac{2}{\pi^3}\left(\frac{a}{\rho_{1n}}\right)^2 \left(\frac{\pi}{6}\right)^{1/6} N^{-5/6}\approx 1, \\
\label{eq:PhiInfPhia}
\langle \Phi^\infty|\Phi^a \rangle_\Omega &\approx 1 - 0.151 N^{-5/2} \approx 1,
%\langle a,n|0,0 \rangle
 %\left[ \int d\rho F^a(\rho)F^0(\rho) \right] \left[ \int d\Omega \Phi^a(\Omega) \Phi^0(\Omega) \right]
% \langle \infty,0|a,n \rangle 
 \end{align}
%\end{widetext}
since $N$ is large and $a/\rho_{1n}$ is small.

\begin{figure}[!htbp]
\centering
\includegraphics[scale=0.6]{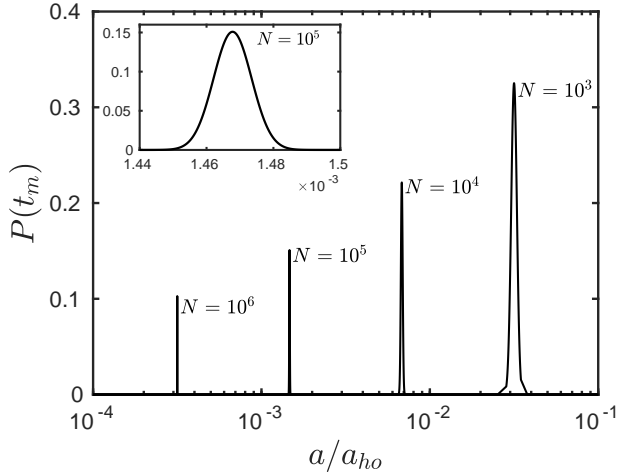}
\caption{\label{fig:TransferProbDiffN} Transfer probability of the BEC versus scattering length $a$ for large $N$. Inset shows a zoom-in profile of $N=10^5$.}
\end{figure}

Next, we convert the discrete sum in Eq.~(\ref{eq:TransitionAmp}) into a continuum integral over the energy and evaluate it at $t=t_m$ around which the maximum transfer occurs. See Appendix~\ref{appendix:TransAmp} for details. The resulting transition amplitude is 
\begin{widetext}
\begin{align}
\label{eq:TransitionAmpLargeN}
  \langle \infty,0|\Psi^a(t_m)\rangle \approx \frac{2 (2d_0)^{1/6}}{(3c_0)^{5/24}\sqrt{3}}N^{1/36}\left(\frac{a}{a_{ho}}\right)^{1/6}\exp\left[ -\left(\left( \frac{2d_0}{\sqrt{3c_0}}\right)^{1/3} N^{13/18} \left(\frac{a}{a_{ho}}\right)^{1/3} -\sqrt{\frac{3N}{2}} \right)^2 \right],
\end{align}
\end{widetext}
where $c_0$ and $d_0$ are defined in Eqs.~(\ref{eq:Vinf}) and ~(\ref{eq:Va}). Plots of $P(t_m)$, calculated in this way, for different $N$ are shown in Fig.~\ref{fig:TransferProbDiffN} as a function of the intermediate scattering length $a$. We see that the estimated maximum transfer for $N=10^3$ is  $\sim33\%$, which is close to what the exact calculation gives. The inset in Fig.~\ref{fig:TransferProbDiffN} shows the sensitivity of the transition probability to the intermediate $a$ for $N=10^5$. The intermediate $a$ should at least be within $0.4\%$ from the optimum to get at least half of the maximum transfer. By maximizing Eq.~(\ref{eq:TransitionAmpLargeN}) with respect to $a$, or by using Eqs.~(\ref{eq:Intermediate_a}), (\ref{eq:rho_0}), and (\ref{eq:rho_inf}), the optimum scattering length is found to be
\begin{equation}
 a=(3/2)^{3/2}\sqrt{3c_0}/(2d_0) N^{-2/3} a_{ho}\approx 3.16 N^{-2/3} a_{ho}.
\end{equation}
And the maximum transfer is
\begin{align}
 {\rm{max}}\left(|\langle \infty,0|\Psi^a(t_m)\rangle|^2\right) & \approx \left|\left(\frac{8}{3} \right)^{1/4} \frac{1}{\left(3c_0\right)^{1/8}} N^{-1/12}\right|^2 \nonumber\\
 & \approx |1.014 N^{-1/12}|^2 \approx 1.028 N^{-1/6}.
\end{align}
To put this into context, for $^{85}$Rb in a trap with frequency $\omega=2\pi \times 10 \,\text{Hz}$, the oscillator length is $a_{ho}=6.51\times10^4\, a_0$. Starting with $N=10^5$ non-interacting atoms in the trap, the two-step process would be optimized for a scattering length of $a^*\approx95.4\,a_0$.

\section{Conclusions and Prospects}

We have presented a protocol designed to implant a nontrivial fraction of the trapped atoms into a resonant BEC.  It remains to be understood what the consequences of this preparation step will be.  It is not clear, for example, what further reorganization of the atoms might be necessary for the gas to resemble an equilibrium resonant BEC.  It is equally unclear at present how three-body losses would differ in the resonant BEC than in a gas of equivalent density.  A useful initial experiment might be to prepare the resonant BEC as proposed here, and compare its dynamics to that of a gas of equal initial density as the resonant BEC, but jumped suddenly to resonance.

This experiment would unfortunately  be clouded by another issue.  Consider, for example, that starting from a non-interacting BEC of $N=10^4$ atoms, our protocol is expected to transfer only one fifth of them to the resonant BEC.  What becomes of the rest?  They are presumably projected onto other quantum mechanical states of the system, each of which has its own dynamics and three-body loss rates.  To address this, it is necessary to formulate a reliable theory of excited states, in our case in the hyperangular degrees of freedom. This pursuit is currently underway. 

\begin{acknowledgements}
 M. W. C. S. and J. L. B. were supported by the JILA NSF Physics Frontier Center, grant number PHY-1734006.
 \end{acknowledgements}

\appendix

\section{Franck-Condon Factors Using the Reflection Formula}
\label{appendix:FC_factors}
Here, we evaluate overlap integrals 
\begin{align}
\label{eq:FaF0}
 \langle F_n^a|F^0 \rangle_\rho &=\int_0^\infty d\rho \,  F_n^a F^0,\\
 \label{eq:FInfFa}
 \langle F^\infty|F_n^a \rangle_\rho &=\int_0^\infty d\rho \,  F^\infty F_n^a. 
\end{align}
Leading contribution to the Franck-Condon factors comes from the overlap of wave functions at the classical turning points, where the wave functions $F_n^a$ are sharply peaked. In between the turning points, the wave functions are highly-oscillating. Yet, we can consider that the projections of $F_n^a$ to $F^0$ and $F^a$ are still localized to the turning points since the latter wave functions are also localized (or close to zero where $F_n^a$ is wildly oscillating). The idea that the Franck-Condon factors can be estimated from properties of the potential near the turning points goes back to the early days of quantum mechanics \cite{Condon_PR28,Winans_28}.  It is widely used in theories of optical and Raman transitions in molecules, and recently to photoassociation of cold atoms as well \cite{Suominen96,Weiner99,Bohn99,Julienne96,Boisseau00}. Out of these types of molecular spectroscopy studies, the reflection formula was developed \cite{Julienne96,Jablonski45}, which we will adapt. 

We first express $F_n^a$ in terms of the energy-normalized wavefunction $F_E$ through
\begin{widetext}
\begin{equation}
 \langle F_n^a | F_{n'}^a \rangle = \int_0^\infty d\rho \, F_n^a  F_{n'}^a = \delta\left(n-n' \right) = \frac{dE_n}{dn} \delta\left(E_n-E_{n'} \right) = \frac{dE_n}{dn} \langle F_E | F_{E'} \rangle,
\end{equation}
\end{widetext}
which leads to $F_n^a=\sqrt{dE_n/dn} F_E$.
Casting $F_E$ into phase-amplitude form, after Milne \cite{Milne30},
\begin{equation}
\label{eq:F_E}
 F_E (k,\rho)\approx \sqrt{\frac{2m}{\pi \hbar^2}} \zeta\left(k\right) \sin\left[\beta(k,\rho) \right],
\end{equation}
where the amplitude $\zeta$ and phase $\beta$ satisfy
\begin{align}
 \left(\frac{\mathrm d^2}{\mathrm d \rho^2} + k^2\left(\rho,E\right) \right) \zeta - \frac{1}{\zeta^3}&=0, \\
 \frac{\mathrm d \zeta}{\mathrm d \rho} -\frac{1}{\beta^2} &=0,
\end{align}
with the wave vector
\begin{equation}
 k(\rho) = \sqrt{\frac{2m}{\hbar^2}\left(E-V(\rho) \right)}.
\end{equation}
The rapid oscillations of $F_E$ in ~(\ref{eq:F_E}) will have negligible effect on the integrals in Eqs.~(\ref{eq:FaF0}) and~(\ref{eq:FInfFa}), where $F_n^a$ is expressed in terms of $F_E$, except when $\rho$ is near a turning point which is also a point of stationary phase. Away from a turning point, it is sufficient to use the WKB approximations for the amplitude and phase:
\begin{align}
 \zeta\left(k\right) &=\frac{1}{\sqrt{k(\rho,E)}}, \\
 \beta\left(k,\rho\right) &= \int_{\rho_t}^{\rho}  d\rho' \, k\left(\rho',E\right) + \frac{\pi}{4}.\\
\end{align}
Near a turning point $\rho_t$, we expand the Milne phase to second order
\begin{align}
 \beta &\approx b_0 + b_1(\rho - \rho_t) + \frac{b_2}{2} (\rho -\rho_t)^2 + ...\\
 b_0&=\frac{\pi}{4}, \\
 b_1 &= \left.\frac{\partial \beta}{\partial \rho} \right|_{\rho=\rho_t} = k(\rho_t,E) =0,\\
 b_2 &= \left.\frac{\partial^2 \beta}{\partial \rho^2} \right|_{\rho=\rho_t} =\left.\frac{\partial k}{\partial \rho} \right|_{\rho_t} = -\frac{m}{\hbar^2} \zeta^2\left(k(\rho_t) \right) \left.\frac{\partial V}{\partial \rho}\right|_{\rho_t}.
 \end{align}
Now, with
\begin{equation}
 F_n^a = \sqrt{\frac{dE_n}{dn}} \sqrt{\frac{2m}{\pi \hbar^2}} \zeta\left(k\right) \sin\left[\beta(k,\rho) \right],
\end{equation}
the integrand $F_n^a F^0$ is sharply localized around $\rho_{1n}$, the classical inner turning point. Thus,
 \begin{widetext}
 \begin{align}
  \langle F_n^a|F^0 \rangle_\rho \approx F^0(\rho_{1n}) \int_0^\infty d\rho \, F_n^a (\rho)= F^0 (\rho_{1n}) \sqrt{\frac{dE_n}{dn}} \sqrt{\frac{2m}{\pi \hbar^2}} \zeta\left(k(\rho_{1n})\right) \int_0^\infty d\rho\, \sin \left[b_0 + \frac{b_2}{2} (\rho - \rho_{1n})^2 \right].
 \end{align}
\end{widetext}
To evaluate the last integral, we use the formula
\begin{equation}
 \int_0^\infty dx \cos\left(x^2\right) = \int_0^\infty dx \sin\left(x^2\right) = \frac{1}{2}\sqrt{\frac{\pi}{2}}.
\end{equation}
Finally, we arrive at
\begin{equation}
\label{eq:FC1n}
 \langle F_n^a|F^0 \rangle_\rho \approx F^0 (\rho_{1n}) \sqrt{\frac{dE_n}{dn}} \sqrt{\frac{1}{|\partial V/ \partial \rho|_{\rho_{1n}}}}.
 \end{equation}
The other overlap factor (\ref{eq:FInfFa}) can be approximated in a similar fashion; it is given by
\begin{equation}
\label{eq:FC2n}
\langle F^\infty|F_n^a \rangle_\rho \approx (-1)^n F^{\infty} (\rho_{2n}) \sqrt{\frac{dE_n}{dn}} \sqrt{\frac{1}{|\partial V/ \partial \rho|_{\rho_{2n}}}},
 \end{equation}
 where the $(-1)^n$ accounts for the sign of the rightmost amplitude around the outer turning point $\rho_{2n}$ of the vibrational state if we set the leftmost amplitude around $\rho_{1n}$ always positive as expressed in Eq.~(\ref{eq:FC1n}).

%DB \section{Weak interaction in the large $N$ limit}
\section{Overlap between LOCV Hyperangular Wave Functions}
\label{appendix:AngularOverlap}
To give a complete picture of the overlap between wave functions, the angular overlaps $\langle \Phi^a|\Phi^0 \rangle_\Omega$ and $\langle \Phi^\infty|\Phi^a \rangle_\Omega$ should also be considered. Real calculation involves $3N-4$ dimensional integrals since this is the size of the hyperangular space. However, here, we only consider the one hyperangle, $\alpha$, that describes the two-body interactions, and the large $N$ case. 

We start with a symmetrized Jastrow-type basis,
\begin{eqnarray}
Y_\nu =  \frac{ \prod_{i<j} \phi_\nu(\rho; \alpha_{ij}) }{ \int d\Omega \sqrt{ \prod_{i<j} \phi_\nu(\rho; \alpha_{ij}) } },
\label{eq:Jastrow}
\end{eqnarray}
where $\alpha_{ij}$ is parametrically related to the coordinate distance between to particles, $r_{ij}$ through $r_{ij}=\sqrt{2}\rho \sin \alpha_{ij}$; the function $\phi_\nu$ satisfies the Bethe-Peierls boundary condition which describes what happens when two particles are close to each other. The other boundary condition is set by treating $|\phi_\nu|^2$ as a pair correlation function such that if two atoms are more than  distance $r_d=\sqrt{2}\rho \sin \alpha_d$ apart, then they become uncorrelated or $|\phi(\alpha_{ij}\geq \alpha_d)|^2=1$. Therefore, within a region bounded by $\alpha_d$, there is on the average only one other atom (out of $N-1$) which can be seen by a fixed atom, or
\begin{equation}
\label{eq:LOCV}
 \frac{4\pi \int_0^{\alpha_d} d\Omega_\alpha \int d\Omega_{N-2}
\prod_{i<j} |\phi_\nu(\rho;\alpha_{ij})|^2}{\int d\Omega_{N-1}
\prod_{i<j} |\phi_\nu(\rho;\alpha_{ij})|^2 } =\frac{1}{N-1},
\end{equation}
where $d\Omega=d\Omega_{N-1}=4\pi d\Omega_\alpha d\Omega_{N-2}$, and $d\Omega_\alpha = \sin^2 \alpha \cos^{3N-7} \alpha\,d\alpha$. If $\alpha_d=\pi/2$, then the right side of ~(\ref{eq:LOCV}) should be one. The full form of the pair correlation function $g_2$ can be written as
\begin{equation}
% \begin{align}
g_2(\alpha) =  \left(4 \pi \int_0^{\pi/2} d\Omega_\alpha \right)
\frac{ \int d\Omega_{N-2} \prod_{i<j} |\phi_{\nu}(\rho;\alpha_{ij})|^2 }
{ \int d\Omega_{N-1} \prod_{i<j} |\phi_{\nu}(\rho;\alpha_{ij})|^2 },
%\end{align}
\end{equation}
which is hard to evaluate. To lowest order, however, it is approximated to be $g_2(\alpha)=|\phi_\nu(\alpha)|^2$. This whole procedure outlined above describes a lowest order constraint variational (LOCV) method in hyperspherical coordinates; details can be found in Ref.\cite{Sze18_PRA}. Given $\rho$ and the scattering length $a$, one can then find $\alpha_d$ and $\phi_v$. The angle $\alpha_d$ becomes extremely small as $N$ increases. Hence $\phi(\rho;\alpha_{ij})$ is one in large region of $\alpha_{ij}$  - this is an approximation that leads to $g_2(\alpha)=|\phi_\nu(\alpha)|^2$. 

In the following derivations, we will also treat all the pair wave functions $\phi(\rho,\alpha_{i'j'})$ equivalent to unity, except one pair namely, $\phi(\rho,\alpha_{12})=\phi(\rho,\alpha)$. So,
\begin{align}
\label{eq:phi0phia}
 \langle \Phi^a|\Phi^0 \rangle_\Omega &\approx \mathcal{N}_0 \mathcal{N}_a \int_0^{\pi/2} d\alpha \, \alpha^2 \phi^a(\rho_{1n};\alpha) \phi^0(\rho_{1n};\alpha) \\
 \label{eq:phiInfphia}
 \langle \Phi^\infty|\Phi^a \rangle_\Omega &\approx \mathcal{N}_{\infty} \mathcal{N}_a\int_0^{\pi/2} d\alpha \, \alpha^2 \phi^\infty(\rho_{2n};\alpha) \phi^a(\rho_{2n};\alpha)
\end{align}
where the $\mathcal{N}$'s are some normalization constants so that $\langle \Phi^0|\Phi^0 \rangle_\Omega=1$, $\langle \Phi^a|\Phi^a \rangle_\Omega=1$, and $\langle \Phi^\infty|\Phi^\infty \rangle_\Omega=1$, and \cite{Sze18_PRA}
\begin{align}
\phi^0(\rho;\alpha)&=1\\
 \phi^a (\rho;\alpha) &\approx  A \left(1 - \frac{a}{\sqrt{2}\rho} \frac{1}{\alpha} \right), \quad \text{if }\, \alpha<\alpha_a \\
 \phi^\infty(\rho;\alpha) &=B\frac{\cos\left(\sqrt{6N\nu_\infty}\alpha\right)}{\alpha} \quad \text{if }\, \alpha<\alpha_\infty,\\
 v_\infty &= c_0 N^{2/3}
\end{align}
The wave functions $\phi^a$ and $\phi^\infty$ identically approach unity for $\alpha>\alpha_a$ and $\alpha>\alpha_\infty$, which are given by 
\begin{align}
 \alpha_a &\approx \left(\frac{\pi}{6}\right)^{1/6} N^{-5/6} \\
 \alpha_\infty &= \left(\frac{2\pi}{27} \right)^{1/6} N^{-5/6}.
\end{align}
Note that $\alpha_a$ and $\alpha_\infty$ are extremely small for large $N$ so that the integrals in Eqs.~(\ref{eq:phi0phia}) and~(\ref{eq:phiInfphia}) are over large part of the $\alpha$-space where $\phi^a$ and $\phi^\infty$ are unity. The constants $A$ and $B$ are determined from the continuity boundary condition at $\alpha_a$ and $\alpha_\infty$:
\begin{align}
 A &\approx 1 + \frac{a}{\sqrt{2}\rho} \frac{1}{\alpha_a}, \\
 B &=\frac{\alpha_\infty}{c_1}=\frac{1}{c_1}\left(\frac{2\pi}{27} \right)^{1/6} N^{-5/6},\\
 c_1 &=\cos \left( \sqrt{6N\nu_\infty}\alpha_\infty\right) \approx -0.942.
\end{align}
We then find
\begin{align}
  \mathcal{N}_0 &=\sqrt{\frac{24}{\pi^3}},\\ 
  \mathcal{N}_a & \approx \sqrt{\frac{24}{\pi^3}} \left[ 1 + \frac{2\sqrt{2}}{\pi^3} \frac{a}{\rho} \alpha_a^2 + \frac{4}{\pi^3} \left(\frac{a}{\rho}\right)^2 \alpha_a +... \right],\\
  \mathcal{N}_\infty & \approx \sqrt{\frac{24}{\pi^3}} \left[ 1 -\frac{12}{\pi^3}\gamma N^{-5/2} +...\right],\\
  \gamma &= \frac{c_2}{2c_1\sqrt{6c_0}}\left(\frac{2\pi}{27}\right)^{1/3} + \frac{1}{2c_1^2} -\frac{1}{3}\approx 0.1997\\
 c_2 &=\sin \left( \sqrt{6N\nu_\infty}\alpha_\infty\right) \approx 0.336.
\end{align}
Finally, after a series of algebraic steps and careful bookkeeping of $N$-scaling of the relevant parameters, we find
\begin{align}
 \langle \phi^0 | \phi^a(\rho_{1n})\rangle_{\alpha} & \approx 1 -\frac{2}{\pi^3}\left(\frac{a}{\rho_{1n}}\right)^2 \alpha_a,\\
 \langle \phi^{\infty} | \phi^a(\rho_{2n})\rangle_{\alpha} &\approx 1 - 0.151 N^{-5/2},
\end{align}
which are our approximations for $ \langle \Phi^a|\Phi^0 \rangle_\Omega$ and $ \langle \Phi^\infty|\Phi^a \rangle_\Omega$, respectively. For large $N$, these quantities are both essentially equal to one.

\section{The Transition Amplitude}
\label{appendix:TransAmp}
We evaluate the transition amplitude at $t=t_m\approx \pi/(2\omega)$ at large $N$. In terms of the Franck-Condon factors derived in Appendix \ref{appendix:FC_factors}, we write the transition amplitude as
\begin{widetext}
 \begin{align}
 \langle \infty,0|\Psi^a(t_m)\rangle &\approx \sum_{n=0}^{\infty} (-1)^n F^0 (\rho_{1n}) F^{\infty} (\rho_{2n}) \frac{dE_n}{dn} \sqrt{\frac{1}{|\partial V/ \partial \rho|_{\rho_{1n}}}} \sqrt{\frac{1}{|\partial V/ \partial \rho|_{\rho_{2n}}}} e^{i\omega_n t_m},
%  \nonumber\\
%  &\approx \int_0^{\infty} dE_n \, F^0 (\rho_{1n}) F^{\infty} (\rho_{2n}) \sqrt{\frac{1}{|\partial V/ \partial \rho|_{\rho_{1n}}}} \sqrt{\frac{1}{|\partial V/ \partial \rho|_{\rho_{2n}}}} e^{iE_n t_m/\hbar},
\end{align}
\end{widetext}
with $\omega_n\approx (2+\Delta_n)n \omega$, where $\Delta_n < 1$ ($\Delta_n \ll 1$ for small $a$). Thus,
\begin{align}
 (-1)^n e^{i\omega_n t_m} &\approx  e^{i(n \pi+\omega_n t_m)} = e^{i2n\pi}=1.
\end{align}
Also, using Eqs.~(\ref{eq:rho1n}) and ~(\ref{eq:rho2n}),
\begin{align}
 \left.\frac{\partial V}{\partial \rho}\right|_{\rho_{1n}} &\approx -3\left(\frac{m}{\hbar^2 d_0 N^{7/2} a}\right)^{1/3} E_n^{4/3},\\
 \left.\frac{\partial V}{\partial \rho}\right|_{\rho_{2n}} &\approx \sqrt{2m \omega^2 E_n}.
\end{align}
Converting the discrete sum into an integral over energy, $\sum_n \rightarrow \int dE$, and using the form of $F^0$ and $F^{\infty}$ in Eqs.~(\ref{eq:F0}) and~(\ref{eq:FInf}), and noting that the resulting integrand is strongly peaked at $E^*\approx\sqrt{3c_0}N^{4/3}\hbar\omega/2\approx1.26N^{4/3} \hbar \omega$ (see Eqs.~(\ref{eq:Emax}) and~(\ref{eq:rho_inf})), we get
\begin{widetext}
\begin{align}
\label{eq:TransitionAmpB}
 \left|\langle \infty,0|\Psi^a(t_m)\rangle\right| & \approx \frac{2(2d_0)^{1/6}}{\sqrt{3\pi}(\sqrt{3c_0})^{11/12}}\left(\frac{a}{a_{ho}}\right)^{1/6}N^{-23/36}\frac{1}{\hbar \omega}\int_0^{\infty} dE \,\exp\left[-\frac{(\rho_{1}(E)- \rho^0)^2}{a_{ho}^2}\right]\exp\left[-\frac{(\rho_{2}(E)- \rho^\infty)^2}{a_{ho}^2}\right],
 \end{align}
\end{widetext}
with $\rho_{1}\approx (\frac{\hbar^2}{m} d_0 N^{7/2} a)^{1/3} E^{-1/3}$ and $\rho_{2}\approx \sqrt{2E/(m\omega^2)}$ from Eqs.~(\ref{eq:rho1n}) and ~(\ref{eq:rho2n}). Now, $F^\infty(\rho_2(E))$ is a peaky function of $E$. We can then use the saddle point approximation to solve the integral in Eq.~(\ref{eq:TransitionAmpB}):
\begin{widetext}
\begin{align}
 \int_0^{\infty} dE \,&\exp\left[-\frac{(\rho_{1}(E)- \rho^0)^2}{a_{ho}^2}\right]\exp\left[-\frac{(\rho_{2}(E)- \rho^\infty)^2}{a_{ho}^2}\right] =\hbar \omega \sqrt{\pi}\frac{\rho^\infty}{a_{ho}}\exp{\left[-\frac{\left(\left(2 d_0 N^{7/2} a_{ho}^4 \frac{a}{\rho^{\infty 2}}\right)^{1/3}-\rho_0\right)^2}{a_{ho}^2} \right]}.
\end{align}
\end{widetext}
Finally, expressing $\rho^0$ and $\rho^\infty$ in terms of $N$,
\begin{widetext}
 \begin{align}
  \left|\langle \infty,0|\Psi^a(t_m)\rangle\right| & \approx \frac{2 (2d_0)^{1/6}}{(3c_0)^{5/24}\sqrt{3}}N^{1/36}\left(\frac{a}{a_{ho}}\right)^{1/6}\exp\left[ -\left(\left( \frac{2d_0}{\sqrt{3c_0}}\right)^{1/3} N^{13/18} \left(\frac{a}{a_{ho}}\right)^{1/3} -\sqrt{\frac{3N}{2}} \right)^2 \right].
  \end{align}
\end{widetext}

\end{document}